\documentclass[11pt]{article}
\usepackage{amsmath,amssymb}
\usepackage{graphicx}
\usepackage{authblk}
\usepackage{hyperref}
\usepackage{booktabs}
\usepackage{adjustbox}
\usepackage{colortbl,xcolor}
\usepackage{tabularx}
\usepackage{listings}
\usepackage{xcolor}
\usepackage{float}

\definecolor{headerblue}{RGB}{79,129,189}

\title{spd-metrics-id: A Python Package for SPD-Aware Distance Metrics in Connectome Fingerprinting and Beyond}
\author[1]{Kaosar Uddin*}
\affil[1]{Department of Mathematics and Statistics, Auburn University, USA}
\affil[*]{Correspondence: mzu0014@auburn.edu}

\date{}

\begin{document}
\maketitle

\begin{abstract}
We present \texttt{spd-metrics-id}, a Python package for computing distances and divergences between symmetric positive-definite (SPD) matrices. 
Unlike traditional toolkits that focus on specific applications, \texttt{spd-metrics-id} provides a unified, extensible, and reproducible framework for SPD distance computation. 
The package supports a wide variety of geometry-aware metrics, including Alpha-$z$ Bures--Wasserstein, Alpha-Procrustes, affine-invariant Riemannian, log-Euclidean, and others, and is accessible both via a command-line interface and a Python API. 
Reproducibility is ensured through Docker images, and Zenodo archiving. 
We illustrate usage through a connectome fingerprinting example, but the package is broadly applicable to covariance analysis, diffusion tensor imaging, and other domains requiring SPD matrix comparison. 
The package is openly available at \url{https://pypi.org/project/spd-metrics-id/}.
\end{abstract}

\textbf{Keywords:} Symmetric positive-definite matrices, Riemannian geometry, Distance metrics, Reproducibility, Connectome analysis

\section{Introduction}
Symmetric positive-definite (SPD) matrices arise in a wide range of scientific and engineering domains, including functional connectivity analysis in neuroimaging \cite{finn2015functional,amico2018,abbas2021geodesic}, covariance estimation \cite{bhatia2009positive,BHATIA2016112}, and diffusion tensor imaging \cite{arsigny2007geometric}. 
Because SPD matrices live on a non-Euclidean manifold, standard similarity measures such as Euclidean distance or Pearson correlation may fail to capture their true geometry \cite{venkatesh2020comparing,albladi2025twssentinovelhybridframework,minh2022alpha,uddin2020comparative,dinh2021alpha}. 
This motivates the need for a unified framework for computing SPD-aware distances and divergences.

We introduce \texttt{spd-metrics-id}, a Python package designed to fill this gap. 
The package provides implementations of multiple well-established SPD metrics in one place, along with practical interfaces for research use: a command-line tool for reproducible workflows and a Python API for integration into analysis pipelines. 
Our goal is to lower the barrier for researchers to apply, compare, and reproduce studies involving SPD-based distances \cite{uddin2025alphazdivergenceunveilsdistinct,kaosaruddin_2025_15891140}.

\section{Background and Related Work}
Existing neuroimaging libraries such as Nilearn and the Brain Connectivity Toolbox \cite{finn2015functional,fornito2016,sporns2018} include utilities for matrix analysis, while geometric toolkits such as \texttt{geomstats} provide general manifold operations \cite{arsigny2007geometric,bhatia2009positive}. 
However, these frameworks do not focus on the specific needs of reproducible SPD distance computation across tasks, parcellations, and data modalities \cite{abbas2023tangent,chiem2022improving}. 
\texttt{spd-metrics-id} complements these resources by offering a lightweight, task-oriented framework that is directly usable in domains like connectome fingerprinting \cite{finn2015functional,amico2018,abbas2021geodesic}, while remaining flexible for broader applications such as brain network modeling \cite{vandenheuvel2019,schaefer2018local,yeo2011organization,van2013wu}.

\section{Features}
The package supports:
\begin{itemize}
    \item A variety of SPD-aware metrics: Alpha-Z Bures--Wasserstein, Alpha-Procrustes, Bures--Wasserstein, Affine-Invariant Riemannian,\\
    Log-Euclidean, Pearson-based, and Euclidean.
    \item Command-line interface (CLI) for reproducible, large-scale runs.
    \item Python API for integration into custom workflows.
    \item Reproducibility via Docker images, Zenodo DOI, and publicly available test datasets.
\end{itemize}

\section{Installation}
The package can be installed from PyPI:
\begin{lstlisting}[language=bash]
pip install spd-metrics-id
\end{lstlisting}

Or from GitHub for development:
\begin{lstlisting}[language=bash]
git clone https://github.com/kaosarUddin/b_f.git
cd spd-metrics-id
python -m venv .venv
source .venv/bin/activate  # macOS/Linux
.venv\Scripts\activate     # Windows
pip install -e .
\end{lstlisting}

\section{Command-Line Usage}
After installation, the \texttt{spd-id} console script is available:
\begin{lstlisting}[language=bash]
spd-id \
  --base-path PATH/TO/DATA \
  --tasks REST LANGUAGE EMOTION \
  --scan-types LR RL \
  --resolutions 100 200 \
  --metric alpha_z \
  --alpha 0.99 \
  --z 1.0 \
  --tau 0.00 \
  --num-subjects 30
\end{lstlisting}

Key arguments are summarized in Table~\ref{tab:args}.

\begin{table}[h!]
\centering
\caption{Key CLI arguments for \texttt{spd-id}.}
\label{tab:args}
\begin{tabular}{ll}
\toprule
Argument & Description \\
\midrule
\texttt{--base-path} & Path to root folder containing subject data \\
\texttt{--tasks} & List of tasks (REST, EMOTION, etc.) \\
\texttt{--scan-types} & Two scan directions to compare (e.g., LR RL) \\
\texttt{--resolutions} & Parcellation sizes (100, 200, 300, ...) \\
\texttt{--metric} & Distance metric: alpha\_z, alpha\_pro, bw, ai, log, pearson, euclid \\
\texttt{--alpha, --z} & Parameters for Alpha-based metrics \\
\texttt{--tau} & SPD regularization (default: 1e-6) \\
\texttt{--num-subjects} & Maximum number of subjects \\
\bottomrule
\end{tabular}
\end{table}

\section{Python API Example}
\begin{lstlisting}[language=Python]
import numpy as np
from spd_metrics_id.io import find_subject_paths, load_matrix
from spd_metrics_id.distance import alpha_z_bw
from spd_metrics_id.id_rate import compute_id_rate

base = "connectomes_100/"
lr_paths = find_subject_paths(base, "REST", "LR", [100], n=30)
rl_paths = find_subject_paths(base, "REST", "RL", [100], n=30)

mats_lr = [load_matrix(p) for p in lr_paths]
mats_rl = [load_matrix(p) for p in rl_paths]

D12 = np.array([[alpha_z_bw(A, B, alpha=0.99, z=1.0) for B in mats_rl] 
for A in mats_lr])
D21 = np.array([[alpha_z_bw(A, B, alpha=0.99, z=1.0) for B in mats_lr] 
for A in mats_rl])

id1 = compute_id_rate(D12)
id2 = compute_id_rate(D21)
print("ID_Rate:", (id1 + id2) / 2)
\end{lstlisting}

\section{Demonstration}
To illustrate usage, we apply the package to a connectome fingerprinting task using 30 Human Connectome Project subjects with 100-region parcellation. 
This demonstration is not intended to establish a new benchmark but to show how the package can be used in practice.

\subsection{Pairwise Distance Structure}
\begin{figure}[H]
\centering
\includegraphics[width=\linewidth]{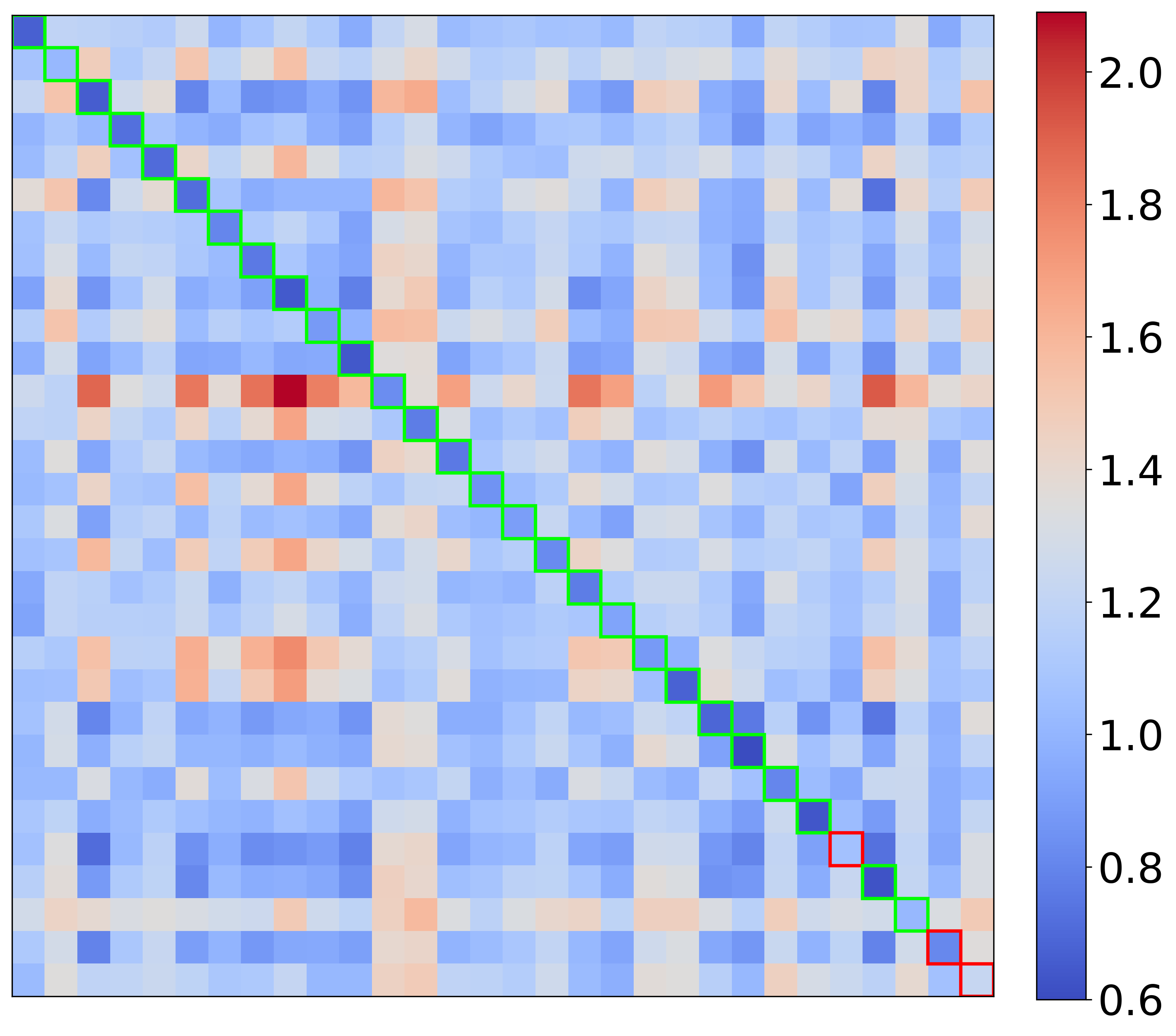}
\caption{Pairwise distance matrix using Alpha-$z$ Bures--Wasserstein divergence ($\alpha=0.99, z=1.0$) for 30 subjects. Correct within-subject matches appear along the diagonal.}
\label{fig:distance_matrix}
\end{figure}

Figure~\ref{fig:distance_matrix} illustrates a pairwise distance matrix 
computed with the Alpha-$z$ Bures--Wasserstein metric. Each row and column 
represents a subject, and the diagonal entries correspond to within-subject 
comparisons. The clear low-distance diagonal (highlighted in green with red markers) 
demonstrates that the package correctly identifies self-connections as more 
similar than connections across different individuals. This example shows 
how \texttt{spd-metrics-id} can visualize and quantify subject-level 
connectome similarity.

Table~\ref{tab:fc-dis2} presents a direct quantitative comparison of 
distances between one anchor subject (Subject~1) and several others 
using multiple metrics. Each metric yields different absolute scales, 
but the relative ranking of within- versus between-subject values 
determines correct identification. For example, although Pearson 
distance for Subject~1 ($0.340$) is numerically smaller than the 
Alpha-$z$ distance ($0.793$), across the broader set of subjects 
Alpha-$z$ consistently assigns the lowest distance to the true self-FC, 
whereas Pearson sometimes fails. This highlights the importance of 
using geometry-aware metrics.

\begin{table}[H]
\centering
\caption{Pairwise distances from the reference subject (Subject~1). 
Each metric provides a numerical distance between Subject~1 and other subjects.
Blue = correct within-subject match; Red = Pearson misleadingly low value.}
\label{tab:fc-dis2}
\renewcommand{\arraystretch}{1.25}
\begin{adjustbox}{max width=\textwidth}
\begin{tabular}{ccccccccc}
\toprule
\rowcolor{headerblue}
{\color{white}\textbf{Subject ID}} &
{\color{white}\textbf{Pair Type}} &
{\color{white}\textbf{Alpha Z BW}} &
{\color{white}\textbf{Alpha Procrustes}} &
{\color{white}\textbf{BW}} &
{\color{white}\textbf{Geodesic}} &
{\color{white}\textbf{Log--Euclidean}} &
{\color{white}\textbf{Pearson}} &
{\color{white}\textbf{Euclidean}} \\
\midrule
1  & $d(1,1)$   & \textcolor{blue}{0.793} & 10.620 & 25.190 & 25.332 & 20.961 & \textcolor{red}{0.340} & 25.936 \\
2  & $d(1,2)$   & 1.043 & 12.879 & 35.845 & 26.596 & 22.004 & 0.368 & 31.292 \\
3  & $d(1,3)$   & 1.316 & 14.575 & 45.087 & 28.152 & 23.822 & 0.511 & 35.428 \\
4  & $d(1,4)$   & 1.244 & 12.792 & 34.762 & 28.171 & 23.005 & 0.465 & 33.613 \\
5  & $d(1,5)$   & 1.176 & 13.252 & 37.312 & 26.857 & 21.955 & 0.472 & 32.034 \\
6  & $d(1,6)$   & 1.317 & 14.403 & 43.328 & 27.278 & 22.611 & 0.524 & 35.055 \\
7  & $d(1,7)$   & 1.216 & 12.460 & 33.840 & 29.475 & 24.063 & 0.444 & 31.707 \\
8  & $d(1,8)$   & 1.166 & 12.732 & 36.147 & 27.475 & 22.629 & 0.447 & 28.846 \\
9  & $d(1,9)$   & 1.140 & 12.351 & 33.684 & 27.995 & 22.985 & 0.414 & 29.561 \\
10 & $d(1,10)$  & 1.282 & 12.898 & 34.726 & 29.739 & 24.302 & 0.427 & 34.475 \\
\midrule
\multicolumn{2}{l}{\textbf{Final match (closest subject)}} &
1 (0.793) & 1 (10.620) & 1 (25.190) & 1 (25.332) & 1 (20.961) & \textcolor{red}{94 (0.272)} & \textcolor{red}{178 (23.754)} \\
\bottomrule
\end{tabular}
\end{adjustbox}
\end{table}

Table~\ref{tab:fc-dis1} further demonstrates this point by comparing 
within-subject distances to the closest other-subject distances for 
three random participants. For Alpha-$z$, the within-subject value is 
always the smallest (blue), leading to correct identification. By 
contrast, Pearson frequently misidentifies another subject as closer 
(red), which undermines its reliability. This example underscores that 
\texttt{spd-metrics-id} not only computes raw distances but also helps 
reveal subtle differences in metric performance.

\begin{table}[H]
\centering
\caption{Within–subject vs.\ closest other-subject distances. 
Blue = Alpha–Z within-subject values; 
Red = Pearson misidentifications.}
\label{tab:fc-dis1}
\renewcommand{\arraystretch}{1.25}
\begin{adjustbox}{max width=\textwidth}
\begin{tabular}{lccccc}
\toprule
\textbf{Subject} & \textbf{AZ: within} & \textbf{AZ: best other} &
\textbf{Pearson: within} & \textbf{Pearson: best other} & \textbf{Closest ID} \\
\midrule
137027 & \textcolor{blue}{0.793} & 0.932 & 0.340 & \textcolor{red}{0.332} & \textcolor{red}{200614} \\
298051 & \textcolor{blue}{0.749} & 0.923 & 0.314 & \textcolor{red}{0.280} & \textcolor{red}{105014} \\
280739 & \textcolor{blue}{0.903} & 0.970 & 0.266 & \textcolor{red}{0.224} & \textcolor{red}{206828} \\
\bottomrule
\end{tabular}
\end{adjustbox}
\end{table}

\section{Reproducibility and Availability}
The package is available at:
\begin{itemize}
    \item PyPI: \url{https://pypi.org/project/spd-metrics-id/}
    \item GitHub: \url{https://github.com/KaosarUddin/b_f}
    \item DockerHub: \url{https://hub.docker.com/r/kaosar148/spd-metrics-id}
    \item Zenodo DOI: \url{https://doi.org/10.5281/zenodo.15891140}
\end{itemize}

\section{Conclusion}
\texttt{spd-metrics-id} provides a unified, reproducible, and extensible framework for computing SPD matrix distances. 
It consolidates multiple geometry-aware metrics into a single package with both CLI and Python API access, lowering the barrier for researchers to apply SPD-based analysis. 
While our demonstration focused on connectome fingerprinting, the package can also be applied to other domains involving SPD matrices, including machine learning, covariance modeling, and medical imaging. 
By making these tools openly available with full reproducibility support, \texttt{spd-metrics-id} aims to serve as a flexible foundation for future research.

\section{Acknowledgements}
The author thanks the open-source Python scientific computing community, whose tools form the foundation for this package.

\nocite{*}
\bibliographystyle{IEEEtran}   
\bibliography{references}  
\end{document}